\documentclass[a4paper]{article}

\def\vpi{\vec{\pi }}
\def\vta{\vec{\tau }}

\def\vsi{\vec{\sigma }}
\def\vga{\vec{\gamma }}
\def\vmu{\vec{\mu }}
\def\hr{\hat{r}}

\def\beu{\hbox{\rm b}}

\def\deu{\hbox{\rm d}}

\def\bd{{\hbox{\rm b}}^d_{m_2,f_2}}

\def\dd{{\hbox{\rm d}}^d_{m_2,f_2}}
\def\bjk{{\hbox{\rm b}}^{d\, \dagger }_{m_1,f_1}}

\def\djk{{\hbox{\rm d}}^{d\, \dagger }_{m_1,f_1}}
\def\ddk{{\hbox{\rm d}}^{d\, \dagger }_{m_2,f_2}}
\def\bmf{{\hbox{\rm b}}^c_{m,f}}

\def\dmk{{\hbox{\rm d}}^{c\, \dagger }_{m,f}}

\def\d{{\rm d}}

\def\bps{\overline{\psi }}
\def\fpi{f_\pi }
\def\mpi{m_{\pi }}
\def\beu{\hbox{\rm b}}
\def\deu{\hbox{\rm d}}

\def\bd{{\hbox{\rm b}}^d_{m_2,f_2}}
\def\dd{{\hbox{\rm d}}^d_{m_2,f_2}}
\def\bjk{{\hbox{\rm b}}^{d\, \dagger }_{m_1,f_1}}

\def\djk{{\hbox{\rm d}}^{d\, \dagger }_{m_1,f_1}}
\def\ddk{{\hbox{\rm d}}^{d\, \dagger }_{m_2,f_2}}
\def\bmf{{\hbox{\rm b}}^c_{m,f}}

\def\dmk{{\hbox{\rm d}}^{c\, \dagger }_{m,f}}
\def\hmf{\chi ^f_m}
\def\img{{\rm i}}
\def\spj{\pmatrix ij_1\\ -(\vsi \hr )j_0\endpmatrix }
\def\spd{\pmatrix i(\vsi \hr ) j_0\\ -j_1\endpmatrix }
\def\spc{\pmatrix (\vsi \hr ) j_0\\ -ij_1\endpmatrix }
\def\spe{\pmatrix j_1\\ -i(\vsi \hr )j_0\endpmatrix }
\def\sel{[\chi ^{f_1\, \dagger }_{m_1} (\vsi \hr )\chi ^{f_2}_{m_2}]}
\def\bra#1{\langle#1\vert}               
\def\ket#1{\vert#1\rangle}               
\def\brik#1#2#3{\langle#1\vert\, #2\,\vert#3\rangle}
\def\slashchar#1{\setbox0=\hbox{$#1$}  
   \dimen0=\wd0     
   \setbox1=\hbox{/} \dimen1=\wd1  
   \ifdim\dimen0>\dimen1   
      \rlap{\hbox to \dimen0{\hfil/\hfil}} 
      #1     
   \else     
      \rlap{\hbox to \dimen1{\hfil$#1$\hfil}} 
      /      
   \fi}      %
\def\vpi{\vec{\pi }}      
\def\vta{\vec{\tau }}     
\def\vsi{\vec{\sigma }}   
\def\vga{\vec{\gamma }}   
\def\vmu{\vec{\mu }}      
\def\hr{\hat{r}}          

\def\bps{\overline{\psi }}

\def\beu{\hbox{\rm b}}
\def\deu{\hbox{\rm d}}

\def\bd{{\hbox{\rm b}}^d_{m_2,f_2}}
\def\ddj{{\hbox{\rm d}}^d_{m_1,f_1}}
\def\dd{{\hbox{\rm d}}^d_{m_2,f_2}}
\def\bjk{{\hbox{\rm b}}^{d\, \dagger }_{m_1,f_1}}

\def\djk{{\hbox{\rm d}}^{d\, \dagger }_{m_1,f_1}}
\def\ddk{{\hbox{\rm d}}^{d\, \dagger }_{m_2,f_2}}
\def\bmf{{\hbox{\rm b}}^c_{m,f}}

\def\dmk{{\hbox{\rm d}}^{c\, \dagger }_{m,f}}

\def\dsmat{\left({\begin{array}{*{20}c}
   {j_0 }  \\
   {\img\left({\vec{\sigma}\hat{r}}\right)j_1}
   \\ \end{array} } \right)}
\def\crmat{\left( {\begin{array}{*{20}c}
   {\left({\vec{\sigma}\hat{r}}\right)j_1}\\
   {\img j_0 }\\
\end{array}}\right)}

\def\spj{\left({\begin{array}{*{20}c} \img j_1\\ -(\vsi \hr )j_0
\\ \end{array} } \right) }
\def\spd{\left({\begin{array}{*{20}c} \img (\vsi \hr ) j_0\\ -j_1
\\ \end{array} } \right) }
\def\spc{\left({\begin{array}{*{20}c} (\vsi \hr ) j_0\\ -\img j_1
\\ \end{array} } \right) }
\def\spe{\left({\begin{array}{*{20}c} j_1\\ -\img(\vsi \hr )j_0
\\ \end{array} } \right) }
\def\sel{[\chi ^{f_1\, \dagger }_{m_1} (\vsi \hr )\chi ^{f_2}_{m_2}]}

\newcommand{\dsp}[1]{{\displaystyle #1}}

\def\fpi{f_\pi }
\def\mpi{m_{\pi }}

\newcommand{\Lagr}{{\cal L}}

\newcommand{\m}[1]{{\bf #1}}

\begin{document}
\thispagestyle{empty}
\vfill
\centerline{\vrule height 1mm width 13cm}
\vskip 0.8truecm
\centerline{\bf THE EXTENDED CHIRAL QUARK MODEL IN A}
\medskip
\centerline{\bf TAMM-DANCOFF INSPIRED APPROXIMATION}
\vskip 0.8truecm
\centerline{\bf Dubravko Horvat \footnote{ e-mail: dubravko.horvat@fer.hr}}
\medskip
\centerline{Department of Physics, Faculty of Electrical Engineering,}
\smallskip
\centerline{University of Zagreb, 10\,000 Zagreb, Croatia}
\vskip 0.5truecm
\smallskip
\centerline{ }
\smallskip
\vskip 0.5truecm
\centerline{{\bf Davor Horvati\' c \footnote{e-mail: davorh@phy.hr}, }
{\bf Boris Podobnik \footnote{e-mail: bp@phy.hr}}
and {\bf Dubravko Tadi\' c \footnote{e-mail: tadic@phy.hr}} }
\medskip
\centerline{Department of Physics, Faculty of Natural Sciences,}
\smallskip
\centerline{University of Zagreb, 10\,000 Zagreb, Croatia}
\vskip 0.8truecm
\centerline{\vrule height 1mm width 13cm}
\smallskip
\vfill
\begin{abstract}
A procedure inspired by the Tamm-Dancoff method is applied to the
chiral quark model which has been extended to include additional
degrees of freedom: a pseudoscalar isoscalar field as well as a
triplet of scalar isovector fields. The simpler, generic $\sigma$
-- model has been used before as a test for the Tamm-Dancoff
inspired approximation (TDIA). The extended chirial quark model is
employed here to investigate possible novel effects of the
additional degrees of freedom as well as to point out the
necessesity to introduce a SU(3) flavour. Model predictions for
the axial-vector coupling constant and for the nucleon magnetic
moment obtained in TDIA are compared with experimental values.
\end{abstract}
\vfill

\newpage
\setcounter{page}{1}

\section{Introduction}
\baselineskip 15pt
\setcounter{equation}{0}

The Tamm-Dancoff method (TDM) \cite{1,2} has been intensively
investigated during the 1950s \cite{3}. It has been revived
recently \cite{4,8,9,10,11}. In some calculations TDM can be a much 
better approximation than perturbation theory~[5].

As chiral bag models \cite{6,7} are simple effective theories of
quark bound states, hadrons, it is not unreasonable to hope that
TDM might be useful in that case too.

The Tamm-Dancoff inspired approximation (TDIA) \cite{12} will be
used here. It will be developed for an extended chiral sigma model
which besides usual modes \cite{6,7} contains additional degrees
of freedom: a pseudoscalar isoscalar field and a triplet of scalar
isovector fields. The U(2) symmetry of the simple model \cite{6,7}
has thus been enlarged to the present U(2)$\times$U(2) \cite{13}.
That allows closer comparison with the SU(3) symmetry based linear
sigma model \cite{14,15}. In such model one has to introduce 9
scalar and 9 pseudoscalar degrees of freedom, thus creating an
U(3)$\times$U(3) symmetry \cite{16,17,18,19}. While in the SU(3)
based case one has 18 mesonic degrees, in our simpler case one
deals with 8 degrees only. The simple $\sigma$-model \cite{12}
used 4 mesonic degrees. Our enlarged U(2)$\times$U(2) model can serve as a
transition to the full SU(3) treatment showing the (non)importance of the scalar
deegres of freedom.   

It will be shown to what extent the new mesonic degrees modify
previously found results \cite{12}.

\section{Tamm-Dancoff inspired approximation}
\setcounter{equation}{0}

Full description of TDIA can be found in ref.~[12]. Here we will give
only some detailes pertinent for the U(2)$\times$U(2) model.  
Working in the Heisenberg picture \cite{20}, all  field operators
are expanded in the free fields \cite{12}. Eventually  one ends
with an infinite set of coupled differential equations instead of
integral ones, which appear in TDM \cite{3,4}. These differential
equations are closely related to the familiar chiral quark model
equations.

The lagrangian containing  the extended linear U(2)$\times$U(2)
sigma model embedded in the bag environment is:

\begin{equation}
\label{2:1} \Lagr = \Lagr _{\psi }\Theta+ \Lagr _{\rm int}\delta
_S+[\Lagr_{\chi }+U(\chi)]\overline{\Theta} 
\end{equation}
where {\setlength\arraycolsep{2pt}
\begin{equation}
\label{2:2}
\begin{array}{rl}
\Lagr_{\psi }&={\displaystyle\frac{\img}2(\bar{\psi }\gamma
^{\mu}\partial _{\mu }\psi -\partial _{\mu }\bar{\psi }\gamma
^{\mu }\psi)}\\ \noalign{\vskip 1ex}
\Lagr_{int}&={\displaystyle\frac{g}2\bar{\psi}(\sigma + \img\vec{\tau
}\vpi\gamma _5)\psi  -\frac{\img g'}{2}\bar{\psi}\gamma_5(\eta+
\img\vec{\tau }\vec{s}\gamma _5)\psi}
  \\ \noalign{\vskip 1ex}
 \Lagr_{\chi}&={\displaystyle\frac12\left(\partial^{\mu }\sigma
\partial _{\mu } \sigma +\partial^{\mu }\vec{\pi }\partial _{\mu }
\vec{\pi }+\partial^{\mu }\eta \partial _{\mu } \eta
+\partial^{\mu }\vec{s }\,\partial _{\mu } \vec{s}\, \right)}\\
\noalign{\vskip 1ex}
U(\chi)&={\displaystyle\frac{\lambda^2}4\left[\sigma ^2+\vec{\pi
}^2+\eta^2+\vec{s}\,^2
  -\nu^2\right]^2 } \\ \noalign{\vskip 1ex}

&{\displaystyle+\frac{\lambda^2\mu^2}2\left[\eta^2+\vec{s}\,^2\right]+f_{\pi}\mpi^2\sigma
 }
\end{array}
\end{equation}}
and $f_{\pi }=0.093\,$GeV. The $\Theta-$function  signalizes that
$\Lagr\, _{\psi }$ is different from zero inside the bag
($r<R_{bag}$). The surface $\delta -$function $\delta _S$ gives
the surface quark - meson interaction, and
$\overline\Theta $ ensures that the potential $U$  and the
$(\sigma ,\vpi, \eta, \vec{s} )$ kinetic-energy terms exist (only) outside the
bag. In the {\sl spherical \/} bag $\Theta $ and
$\overline{\Theta}$ become $\theta (R_{bag}-r)$ and $\theta
(r-R_{bag})$,  respectively. In the product symmetry U(2)$\times$U(2) the
coupling constants $g$ and $g'$ can have different values. In actual
calculations we were forced to break this symmetry thus introducing an
assortment of $g$'s whose values are connected trough boundary conditions (see
(2.13) below). The self-interaction potential $U$
contains the  symmetry-breaking $(SB)$ term $c\sigma (x)
\equiv-f_{\pi }m^2_{\pi }\sigma (x)$. The other parameters are
fixed by particle mass spectrum, i.e. {\setlength\arraycolsep{2pt}
\begin{equation}
\label{2:3}
\begin{array}{rl}
  m_{\sigma}^2&=-\lambda^2\nu^2+3\lambda^2 f_\pi^2 \\\noalign{\vskip 1ex}
  \mpi^2&=-\lambda^2\nu^2+\lambda^2 f_\pi^2 \\\noalign{\vskip 1ex}
  m_{\eta}^2 &=m_{s}^2 =\lambda^2\mu^2-\lambda^2\nu^2+\lambda^2 f_\pi^2 \\
\end{array}
\end{equation}}
by the PCAC and by the usual potential minimum conditions [14,15].

The effective empirical quantum field theory (\ref{2:1}) describes
quark dynamics, as an approximant for the underlying, fundamental
and exact QCD.

The field operators $\psi$, $\vec{\pi}$, $\sigma$, $\eta$ and $\vec{s}$ are
expanded
in terms of the free field creation (annihilation) operators. For the quark field, for
example,
one introduces {\setlength\arraycolsep{2pt}
\begin{equation}
\label{2:4}
\begin{array}{rl} \psi_{f}^c(x)=&\phi_m^f\beu
^c_{m,f}+\tilde{\phi}_m^f \deu ^{c\dagger }_{m,f}\\
\noalign{\vskip 1ex} &+\chi_{m_1m_2m_3}^{fgh}(x)\beu_{m_1,f}^c
\deu_{m_2,g}^{e\dagger }\beu_{m_3,h}^e+\dots
\end{array}
\end{equation}}

Many complex operator combinations, besides the one shown, are posible. However
in our approximation we use only first two terms. 
Here $c$ is a quark colour and $f$ is a quark flavour, whereas $m$
is the spin projection. $ \beu ^c_{m,f} $ and  $\deu ^c_{m,f} $
are quark and antiquark annihilation operators, respectively. This
infinite expansion is truncated leading to a physically motivated
finite basis, which defines the  Tamm-Dancoff inspired
approximation (TDIA).

The truncation of the $\psi$ field (\ref{2:4}) as well as the
corresponding ans\"atze for the $\vec{\pi}$, $\sigma$, $\eta$
 and $\vec{s}$ fields
lead to the following system of the coupled nonlinear
Euler-Lagrange equations
\begin{eqnarray}
\label{2:5}
\img\gamma_\mu\partial^\mu\psi=0 & (r<R_{\rm Bag}),
\nonumber
\\ \nonumber
\partial_\mu\partial^\mu\sigma-\lambda^2\sigma\left(\sigma^2+\vpi^2+\eta^2
+\vec{s}^{\,2}-\nu^2\right)-\fpi\mpi^2=0&(r>R_{\rm Bag}),\\
\nonumber
\partial_\mu\partial^\mu\pi^a-\lambda^2\pi^a\left(\sigma^2+\vpi^2+\eta^2
+\vec{s}^{\,2}-\nu^2\right)=0&(r>R_{\rm Bag}),\\ \nonumber
\partial_\mu\partial^\mu\eta-\lambda^2\eta\left(\sigma^2+\vpi^2+\eta^2
+\vec{s}^{\,2}-\nu^2+\mu^2\right)=0&(r>R_{\rm Bag}),\\
\nonumber
\partial_\mu\partial^\mu s^a-\lambda^2 s^a\left(\sigma^2+\vpi^2+\eta^2
+\vec{s}^{\,2}-\nu^2+\mu^2\right)=0&(r>R_{\rm Bag}),\nonumber
\\
{\displaystyle \img n_\mu\gamma^\mu\psi +g_\sigma \sigma\psi+g_\pi \img
\vta\vpi\gamma_5\psi-g_\eta \img \gamma_5\eta\psi+g_s\vta\vec{s}\psi=0} &
 (r=R_{\rm Bag}), \\
\nonumber
\left(\partial^\mu\sigma\right)\hat{n}_\mu={\displaystyle -
\frac{g_\sigma}{2}}\bps\psi & (r=R_{\rm Bag}),\\ \nonumber
\left(\partial^\mu\pi^a\right)\hat{n}_\mu={\displaystyle -
\frac{g_\pi}{2}} \bps \img\vta\gamma_5\psi & (r=R_{\rm Bag}),\\
\nonumber
\left(\partial^\mu\eta\right)\hat{n}_\mu={\displaystyle\frac{g_\eta}{2}}\bps
\img\gamma_5\psi & (r=R_{\rm Bag}),\\ \nonumber \left(\partial^\mu
s^a\right)\hat{n}_\mu={\displaystyle - \frac{g_s}{2}}\bps\vta\psi
& (r=R_{\rm Bag}) \nonumber
\end{eqnarray}
In TDIA sense one keeps just the terms in the field expansion
which are needed to obtain a nontrivial coupled system of
differential equations.

The leading approximation follows if in the equations (\ref{2:5})
one keeps just two first terms in the expansion (\ref{2:4}).

The result is then sandwiched between the initial quark or
antiquark states
\begin{equation}
\label{2:6}
\begin{array}{ll}
\bra{f}=\bra{\bar{q}_{p,r}^{a}}=\bra{0}{\rm d}_{p,r}^{a}\quad &
\ket{i}=\ket{0}\quad {\rm or}
\\ \noalign{\vskip 2ex}
\bra{f}=\bra{0}\quad &
\ket{i}=\ket{q_{p,r}^{a}}={\rm b}_{p,r}^{a\dag}\ket{0}
\\
\end{array}
\end{equation}
and the vacuum, leading to the terms such as
\begin{equation}\label{2:7}
 \brik{0}{\gamma_{\mu}{\partial}^{\mu}\psi}{q_{f,m}^c}=\gamma_{\mu}
\partial^{\mu}\phi_m^f(x).
\end{equation}
The Dirac equation for the free quark inside the bag [21],
accordingly with (\ref{2:4}) leads to the following approximate
TDIA ansatz for the quark field
\begin{equation}
\label{2:8}
\psi^c_f(x)=\frac{N}{\sqrt{4\pi}}
\left({\begin{array}{*{20}c}
   {j_0 }\\
   {\img\left({\vec{\sigma}\hat{r}}\right)j_1}\\
\end{array} }\right)
\chi^f_m \beu ^c_{m,f}+\frac{N}{\sqrt{4\pi}}
\left( {\begin{array}{*{20}c}
   {\left({\vec{\sigma}\hat{r}}\right)j_1}\\
   {\img  j_0 }\\
\end{array}}\right)
\chi^f_m \deu ^{c\dagger }_{m,f}
\end{equation}
Here the quantities $j_{0,1} (r)$ are spherical Bessel
functions of the order (0,1) and $\chi _m^f$ is the quark
isospinor ($\tilde{\chi }^f$) - spinor ($\chi _m$) product.

%

The boundary conditions (\ref{2:5}) can be
satisfied with the first two terms in the expansion (\ref{2:4}) if
the corresponding $\vec{\pi}$, $\sigma$, $\eta$ and $\vec{s}$
field expansion contains terms such as
\begin{equation}\label{2:9}
 {\beu_{m,f}^c}^{\dagger}\beu_{m',f'}^c.
\end{equation}

All terms in (\ref{2:5}), either the bispinor ones
($\overline{\psi}\Gamma \psi$) or the meson ones ($\vec{\pi},\
\sigma$) must contain the same number and the same kind of the
creation (annihilation) quark operators. Thus the TDIA ans\"atze
given in terms of {\sl chiral-\/}quark operators are [12]:
{\setlength\arraycolsep{2pt}
\begin{equation}
\label{2:10}
\begin{array}{rl}
\sigma (r)&=
{\displaystyle \sigma _s(r)\cdot (\beu ^{c\dagger }_{m,f}\beu^{c}_{m,f}
           + \deu ^{c\dagger }_{m,f}\deu ^{c }_{m,f})-\fpi} \\ \noalign{\vskip 1ex}
& \\
\pi^a(r)&=
{\displaystyle \pi _s(r)(\beu ^{c\dagger }_{m,f}\deu^{c\dagger}_{m',f'}+
          \deu ^c_{m,f}\beu ^c_{m',f'}) \cdot [\chi^{\dagger}_{m,f}\tau ^a
          \chi _{m',f'}]} \\ \noalign{\vskip 1ex}
& {\displaystyle +\pi _p(r)(\beu^{c\dagger}_{m,f}\beu ^c_{m',f'}-
          \deu ^{c\dagger }_{m',f'}\deu^c_{m,f}) \cdot [\chi ^{\dagger }_{m,f}
          (\vsi \hr )\tau ^a\chi_{m',f'}]} \\ \noalign{\vskip 1ex}
& \\
\eta(r)&=
{\displaystyle \eta_s(r)(\beu ^{c\dagger }_{m,f}\deu^{c\dagger}_{m',f'}+
         \deu ^c_{m,f}\beu ^c_{m',f'})} \\ \noalign{\vskip 1ex}
& {\displaystyle +\eta_p(r)(\beu ^{c\dagger}_{m,f}\beu ^c_{m',f'}-
         \deu ^{c\dagger}_{m',f'}\deu ^c_{m,f}) \cdot
         [\chi ^{\dagger }_{m,f}(\vsi \hr)\chi _{m',f'}]}\\ \noalign{\vskip 1ex}
& \\
s^{a}(r)&=
{\displaystyle s_s(r)\cdot (\beu ^{c\dagger}_{m,f} \beu ^{c}_{m,f}+
          \deu ^{c\dagger }_{m,f}\deu ^{c }_{m,f})\cdot
          [\chi ^{\dagger}_{m,f}\tau ^a \chi _{m',f'}]}
\end{array}
\end{equation}}
The Ans\"atze obey spin-isospin and Lorentz properties of
corresponding particles. They were inspired by the valence 
quark content of mesons and they correctly match the quark 
field approximation (2.8).

At this level of TDIA expansion only the quark operators are
important. The boson operators can be introduced later on or one
can assume that the theory (2.1, 2.2) contains the fermions only.
Then the terms like $\sigma^2,\ \vec{\pi}^2$ etc. describe various
nonlinear interactions among fermions (quarks) which have to be
coupled in scalar (pseudoscalar) combinations. Such models
(theories) \cite{7} would be effective nonrenormalizable field
theories.

In the following the terms {\it meson\/}, {\it pion\/}, {\it
sigma\/}, {\it eta\/} or {\it s\/} are used in that generalized
sense refering to expressions like (\ref{2:10}).

The expansions (\ref{2:10}) for bosonic quantities appear quite
naturally. They have been encountered in the past applications of
the Tamm-Dancoff procedure, as for example in the formula (4.6) of
ref. \cite{11}. As the operators $\beu$ and $\deu$ have the
opposite parity [22], the TDIA conserves parity throughout.

The boundary conditions in (\ref{2:4}) are now specified using the
ans\"{a}tze (\ref{2:8}) and (\ref{2:10}). When they are
sandwiched between the final and initial states as given in
(\ref{2:6}), one finds

{\setlength\arraycolsep{2pt}
\begin{equation}
\label{2:11}
\begin{array}{rl}
\dsp{{\frac{\partial }{{\partial r}}\sigma _s(r)}\Big|_{r = R}}&
{\displaystyle=-\frac{g_\sigma N^2}{8\pi}\left[j_0^2(\omega)-j_1^2(\omega)\right]}\\
\noalign{\vskip 1ex}
\dsp{{\frac{\partial }{{\partial r}}\pi_s(r)}\Big|_{r = R}}&
{\displaystyle=+\frac{g_{\pi_s}N^2}{8\pi}\left[j_0^2(\omega)+j_1^2(\omega)\right]}\\
\noalign{\vskip 1ex}
\dsp{{\frac{\partial }{{\partial r}}\pi_p(r)}\Big|_{r = R}}&
{\displaystyle=+\frac{g_{\pi_p} N^2}{4\pi}j_0(\omega)j_1(\omega)}\\
\noalign{\vskip 1ex}
\dsp{{\frac{\partial }{{\partial r}}\eta_s(r)}\Big|_{r = R}}&
{\displaystyle=-\frac{g_{\eta_s}N^2}{8\pi}\left[j_0^2(\omega)+j_1^2(\omega)\right]}\\
\noalign{\vskip 1ex}
\dsp{{\frac{\partial }{{\partial r}}\eta_p(r)}\Big|_{r = R}}&
{\displaystyle=-\frac{g_{\eta_p}N^2}{4\pi}j_0(\omega)j_1(\omega)}\\
\noalign{\vskip 1ex}
\dsp{{\frac{\partial }{{\partial r}} s_s(r)}\Big|_{r = R}}&
{\displaystyle=-\frac{g_s N^2}{8\pi}\left[j_0^2(\omega)-j_1^2(\omega)\right]}
\end{array}
\end{equation}}
Here $\omega$'s are eigenfrequencies determined by boundary conditions (2.12).
The normalisation N is defined by (4.1) below. 
From $\Lagr_{int}$ one can derive
{\setlength\arraycolsep{2pt}
\begin{equation}
\label{2:12}
\begin{array}{rl}
\dsp{\psi (r)\Big\vert _{r=R_{\rm bag}}} &=\dsp{\img g_{\sigma }\sigma
(r)(\vga \hr )\psi (r) \Big\vert _{r=R_{\rm bag}}- g_{\pi }\vta \vpi
(r)(\vga \hr )\gamma _5\psi (r)\Big\vert _{r=R_{\rm bag}}+}
\\\noalign{\vskip 1ex} &\dsp{+g_\eta \eta(r) (\vga \hr) \gamma_5
\psi(r)\Big\vert _{r=R_{\rm bag}}+ \img g_s \vta \vec{s} (r) (\vga \hr)
\psi(r)\Big\vert _{r=R_{\rm bag}}}
\end{array}
\end{equation}}
It takes the following form:


{\setlength\arraycolsep{2pt}
\begin{eqnarray}
& &\dsmat \chi^f_m \beu ^c_{m,f}+\crmat \hmf \dmk = \nonumber \\
&-&g_{\pi /p}\pi _p(R)\spc \hmf \bjk (\vta \cdot \vta )\bd \sel
\bmf \nonumber \\ &+&g_{\pi /p}\pi _p(R)\spc \hmf \ddk (\vta \cdot
\vta )\ddj \sel \bmf \nonumber \\ &-&g_{\pi /p}\pi _p(R)\spe \hmf
\bjk (\vta \cdot \vta)\bd \sel \dmk \nonumber \\ &+&g_{\pi /p}\pi
_p(R)\spe \hmf \ddk (\vta \cdot\vta )\ddj \sel \dmk \nonumber \\
&+&\img g_{\sigma }\sigma _s(R)\spj \hmf \bjk \bd \bmf \nonumber \\
&+&\img g_{\sigma }\sigma _s(R)\spj \hmf \djk \dd \bmf\nonumber \\
&-&\img g_{\sigma }\fpi \spj \hmf  \bmf \nonumber \\ &+&\img g_{\sigma
}\sigma_s(R)\spd \hmf \bjk \bd \dmk \nonumber \\ &+&\img g_{\sigma
}\sigma _s(R)\spd\hmf \djk \dd \dmk \nonumber \\ &-&\img g_{\sigma
}\fpi \spd \hmf \dmk \nonumber \\ &-&g_{\pi/s}\pi _s(R)\spc \hmf
\bjk (\vta \cdot \vta )\ddk \bmf \nonumber \\ &-&g_{\pi/s}\pi
_s(R)\spc \hmf \ddj (\vta \cdot \vta )\bd \bmf \nonumber \\
&-&g_{\pi/s}\pi _s(R)\spe \hmf \bjk (\vta \cdot \vta )\ddk \dmk
\nonumber \\ &-&g_{\pi/s}\pi _s(R)\spe \hmf \ddj (\vta \cdot \vta
)\bd \dmk + \ldots \nonumber \\
&+&g_{\eta /p}\eta _p(R)\spc \hmf \bjk \bd \sel \bmf \nonumber \\
&-&g_{\eta /p}\eta _p(R)\spc \hmf \ddk \ddj \sel \bmf \nonumber \\
&+&g_{\eta/p}\eta _p(R)\spe \hmf \bjk \bd \sel \dmk \nonumber \\
&-&g_{\eta /p}\eta_p(R)\spe \hmf \ddk \ddj \sel \dmk \nonumber \\
&+&\img g_{s}s_s(R)\spj \hmf \bjk(\vta \cdot \vta ) \bd \bmf \nonumber
\\ &+&\img g_{s}s_s(R)\spj \hmf \djk (\vta\cdot \vta )\dd
\bmf\nonumber \\
&+&\img g_{s}s_s(R)\spd \hmf \bjk(\vta \cdot \vta ) \bd \dmk \nonumber
\\ &+&\img g_{s}s_s(R)\spd \hmf \djk(\vta \cdot \vta ) \dd \dmk
\nonumber \\
&+&g_{\eta /s}\eta _s(R)\spc \hmf \bjk \ddk \bmf \nonumber \\
&+&g_{\eta/s}\eta _s(R)\spc \hmf \ddj \bd \bmf \nonumber \\
&+&g_{\eta /s}\eta_s(R)\spe \hmf \bjk \ddk \dmk \nonumber \\
&+&g_{\eta /s}\eta _s(R)\spe \hmf \ddj \bd \dmk  \nonumber
\end{eqnarray}}


After sandwiching \cite{12} one obtains
\begin{equation}
\begin{array}{l}
\label{2:13}
\dsp{j_0(\omega)[g_{\sigma }(\fpi -\sigma _s(R))+3g_s s_s(R)]
-j_1(\omega)(1-3g_{\pi /p}\pi _p(R)+g_{\eta /p} \eta_p(R))=0,}\\
\\
\dsp{j_0(\omega)(1+3g_{\pi/p}\pi _p(R)-g_{\eta /p} \eta_p(R))
-j_1(\omega)[g_{\sigma }(\fpi -\sigma _s(R))-3g_s s_s(R)]=0, }\\
\\
\dsp{j_0(\omega)- j_1(\omega)(g_{\sigma }\fpi +3g_{\pi
/s}\pi_s(R)-g_{\eta /s} \eta_s(R))=0,}\\ \\
\dsp{j_0(\omega)(g_{\sigma }\fpi -3 g_{\pi /s}\pi _s(R)+g_{\eta
/s} \eta_s(R)) -j_1(\omega)=0};\qquad R=R_{\rm Bag} \\
\end{array}
\end{equation}
In all the above relations flavour and angular-momentum dependent
strong coupling constants $g_{\pi/p}$, $g_{\eta/p}$, $g_{\eta/s}$
etc. appear. This reflects chiral symmetry breaking, which appears naturally
when the non-linear system (2.2) is solved using the Ans\"atze
(\ref{2:8})-(\ref{2:10}).

In order to extract the equations  for the $s-$ and $p-$ wave
components from the equations (\ref{2:5}) they are "sandwiched"
between the final state $\bra{f}=\bra{q^c_{f,t}}=\bra{0}\beu
^c_{f,t}$ and the initial state $\ket{i}=\ket{q^c_{i,u}}= \beu ^{c
\dagger}_{i,u}\ket{0}$. This choice yields the equations for
$\sigma$, $\pi$, $\eta$ and $s$ fields
{\setlength\arraycolsep{2pt}
$$
\begin{array}{rl}
\dsp{\left[\frac{\d^2}{\d
r^2}+\frac2{r}\frac{\d}{\d r}\right]\sigma_s(r)}&= \dsp{
\fpi\lambda^2\left(\fpi^2-\nu^2\right)+} \\\noalign{\vskip 1ex}
&\dsp{+\lambda ^2\left[\sigma _s(r)-\fpi \right]\left[(\sigma
_s(r)-\fpi )^2+ 3\pi _p(r)+\eta_p^2+3s_s^2-\nu ^2 \right]}
\end{array}
$$ } $$
\left[\frac{\d^2}{\d r^2}+\frac2{r}\frac{\d}{\d r}\right]\pi_s(r)
=\lambda ^2\pi_s(r)\left[\fpi^2+ 36\pi_s(r)+12\eta_s^2-\nu ^2
\right]
$$ $$
\left[\frac{\d^2}{\d r^2}+\frac2{r}\frac{\d}{\d
r}-\frac{2}{r^2}\right]\pi_p(r)=\lambda ^2\pi_p\left[(\sigma
_s(r)-\fpi )^2+ 3\pi _p(r)+\eta_p^2+3s_s^2-\nu ^2 \right]
$$ 
\begin{equation}
\label{2:14} \left[\frac{\d^2}{\d r^2}+\frac2{r}\frac{\d}{\d
r}\right]\eta_s(r) =\lambda ^2\eta_s(r)\left[\fpi^2+
36\pi_s(r)+12\eta_s^2-\nu ^2+\mu^2 \right]
\end{equation} $$
\left[\frac{\d^2}{\d r^2}+\frac2{r}\frac{\d}{\d
r}-\frac{2}{r^2}\right]\eta_p(r)=\lambda^2\eta_p\left[(\sigma
_s(r)-\fpi )^2+ 3\pi _p(r)+\eta_p^2+3s_s^2-\nu ^2+\mu^2 \right]
$$ $$
\left[\frac{\d^2}{\d r^2}+\frac2{r}\frac{\d}{\d
r}\right]s_s(r)=\lambda^2 s_s\left[(\sigma _s(r)-\fpi )^2+ 3\pi
_p(r)+\eta_p^2+3s_s^2-\nu ^2+\mu^2 \right]
$$
The problem is to find a set of solutions of the differential
equations (\ref{2:14}), \{~$\sigma_s$, $\pi_s$,
$\pi_p$, $\eta_s$, $\eta_p$, $s_s$~\}, which satisfy the {\sl mathematical}
boundary conditions (\ref{2:11}). These solutions must be
compatible with (\ref{2:13}) which is independent of $r$ so one
has a strongly correlated algebraic system (\ref{2:13}) and the
system of differential equations.

The parameters $(\lambda , \nu, \mu )$ which enter $\Lagr\,$ (2.2)
are restricted by the symmetry - breaking behaviour of the theory.
Usually  [6, 23],  the $\sigma $ particle is considered to be
a 1.2 GeV resonance, whereas the meson "masses" are parameter
which, for simplicity (and  lack of knowledge), are assigned the
values of the physical masses. We have used rounded up values
$m_\pi=0.140\,$GeV, $m_\eta=0.980\,$GeV and $m_s=0.980\,$GeV. Here
we have tentatively identified $s$ meson with $a_0(980)$
$\left[I^G(J^{PC}=1^-\left(0^{++}\right)\right]$ and $\eta$ with
$\eta'(958)$. In the numerics we have also used an alternative
value $m_\sigma=0.450\,$GeV [24]. In the present application,
these "experimental" values  have been used, although $m_\sigma$,
$m_{\pi}$ etc. can, in principle, be considered as additional
parameters.

The usage of the bag-model has to some extent decoupled the
equation for the quark expansion functions $\phi_m^f$,
$\tilde{\phi}_m^f$ etc. (\ref{2:4}) from the rest. It communicates
with the $\pi_n(r),\eta_n(r)$ (n=s, p) and $\sigma(r),s(r)$
functions only through algebraic relations (\ref{2:13}).

Higher order terms in the expansion, such as the third term in
(\ref{2:4}) for example, would enlarge the system of the coupled
equations. As in TDM the whole system would be coupled sector by
sector. That would be governed by the number of creation
(annihilation) operators and by some additional $\ket{i}$
($\bra{f}$) states besides (\ref{2:6}) ones. The end results would
be analogous to the relations among different sectors in the Fock
space in TDM, as one should expect from its reversed picture.

\section{Coupling constants and magnetic moment}
\setcounter{equation}{0}

The results obtained in the leading order of TDIA are used to
calculate the proton magnetic moment and the axial vector coupling
constants.

The magnetic moment operator is
\begin{equation}
\vmu (\vec r)=\frac12({\vec r}\times {\vec j}_{EM}(\vec r)).
\end{equation}
Here
\begin{equation}
j^{\mu }_{EM}(r)=\bps (r)\gamma ^{\mu }Q\psi (r)+\epsilon _{3ij}\pi _i(r)
\partial ^{\mu }\pi _j (r)+\epsilon _{3ij} s _i(r)
\partial ^{\mu } s_j (r)
\end{equation}
\begin{equation}
Q=\frac23\cdot \frac{1+\tau _3}2-\frac13\cdot \frac{1-\tau _3}2.
\end{equation}

The quark contribution to the magnetic moment is
\begin{equation}
\mu ^{(q)}=\frac23\cdot \frac{R}{\omega ^4}\cdot
\frac{(\omega /2)-(3/8) \sin 2\omega +(\omega /4)\cos 2\omega }
{j^2_0(\omega )+j^2_1(\omega )-2 j_0(\omega ) j_1(\omega )/\omega}.
\end{equation}
The meson contribution is
\begin{equation}
\mu ^{(M)}_p=
\frac{16\pi }3\cdot \frac {11}3
\int ^{\infty }_{R_{\rm bag}}r^2\, \d r \left(\pi _p(r)^2+s_s(r)^2\right).
\end{equation}
The proton magnetic moment is given by
\begin{equation}
\mu _p=\mu ^{(q)}+\mu _p^{(M)}.
\end{equation}

The axial vector current
\begin{equation}
J_A^\mu=\bps\gamma^\mu\gamma_5\frac{\vta}{2}\psi+\sigma\partial^\mu\vpi-
\vpi\partial^\mu\sigma+\eta\partial^\mu\vec{s}-\vec{s}\partial^\mu\eta
\end{equation}
leads to the quark contribution:
\begin{equation}
\begin{array}{rl}
g^{(q)}_A&{\displaystyle =\bra{p\uparrow }\int \d^3\vec r \bps ({\vec r})\gamma ^3\gamma ^5
\frac {\tau ^3}2 \psi ({\vec r})\ket{n \uparrow }}\\
&{\displaystyle =\frac53\cdot \frac13\cdot
\frac{j^2_0(\omega )+j^2_1(\omega )}
{j^2_0(\omega )+j^2_1(\omega )-2 j_0(\omega ) j_1(\omega )/\omega}.}
\end{array}
\end{equation}
and to the meson contribution:
\begin{equation}
g^{(M)}_A=\frac53 \cdot \frac{4\pi }3\cdot
\int_{R_{\rm bag}}^{\infty }\d r\,
r^2\big[(\sigma _s(r)-\fpi)\big[\pi '_p(r)+\frac{2\pi _p(r)}r\big]
-\pi _p(r)\sigma ' _s(r)\big].
\end{equation}
Finally:
\begin{equation}
g_A^{(p)}=g_A^{(q)}+g_A^{(M)}.
\end{equation}

The isoscalar axial vector  current

\begin{equation}
{J_A^0}^\mu=\bps\gamma^\mu\gamma_5\psi+\sigma\partial^\mu\eta-\eta\partial^\mu\sigma+
\vec{s}\partial^\mu\vpi-\vpi\partial^\mu\vec{s}
\end{equation}

leads to quark contribution to the isoscalar coupling constant
\begin{equation}
\begin{array}{rl}
{g^{0}_A}^{(q)}&{\displaystyle =\frac13\cdot
\frac{j^2_0(\omega )+j^2_1(\omega )}
{j^2_0(\omega )+j^2_1(\omega )-2 j_0(\omega ) j_1(\omega )/\omega}.}
\end{array}
\end{equation}
The corresponding meson contribution is:
\begin{equation}
\begin{array}{rl}
{g^0_A}^{(M)}&={ \displaystyle\frac{4\pi }3\cdot
\int_{R_{\rm bag}}^{\infty }\d r\,
r^2\big[
\eta_p(r)\sigma_s^\prime(r)-(\sigma_s(r)-\fpi)
\left(\eta_p^\prime(r)+\frac{2 \eta_p(r)}{r}\right)+}\\
&{\displaystyle +5 s_s \left(\pi_p^\prime(r)+\frac{2 \pi_p(r)}{r}\right)
-5 \pi_p(r) s_s^\prime(r)}
\big].
\end{array}
\end{equation}
Finally:
\begin{equation}
{g_A^{0}}^{(p)}={g^0_A}^{(q)}+{g^0_A}^{(M)}.
\end{equation}

\section{Numerical procedure}
\setcounter{equation}{0}

Numerics will be illustrated here for a non-linear system of
coupled ordinary differential equations which have been derived
in sect. 2.

This  system  determines fermion and boson radial
functions appearing in the ans\"atze, (\ref{2:8}) and (\ref{2:10}).
The boson radial functions had to satify the eqs. (\ref{2:11}), (\ref{2:13}) and
(\ref{2:14}).

In (\ref{2:13}) the normalization constant $N$ can
be expressed in terms of spherical Bessel functions and quark 
eigenfrequencies
$\omega $:
\begin{equation}
\label{4:1}
N^2=\frac1{R^3}\left[j^2_0(\omega)+j^2_1(\omega)-
\frac{2j_0(\omega)j_1(\omega)}{\omega }\right]^{-1}.
\end{equation}

The radial parts of the quark wave functions appearing in (\ref{2:8})
are spherical Bessel functions $j_{\ell}(\omega r/R)$ for any spherical bag with
radius $R$. At the bag boundary, where $r=R$, these functions have to satisfy
the relations (\ref{2:13}) which combine the quark
frequency $\omega$ with
the coupling constants $g_{\sigma },\ g_{\pi },\ \fpi \ $ etc.

The linear $\sigma $-model parameters satisfy the following
relations derived from the symmetry breaking pattern (see sect. 2)
\cite{6}:
\begin{equation}
\label{4:2}
\lambda ^2=\frac {m^2_{\sigma }-\mpi ^2}{2\fpi ^2},\qquad
\nu ^2=\fpi ^2-\frac {\mpi ^2}{\lambda ^2},\qquad
\mu=2 \fpi^2 \frac{m^2_{\eta }-
3\mpi ^2}{m^2_{\sigma }-\mpi ^2}.
\end{equation}

The $\sigma $ meson is expected to have a mass of about 1 GeV
[23]. Thus the {\sl parameter} masses $m_{\sigma }$, $m_{\pi
}$, $m_{\eta }$ and $m_{s}$ are selected to be 1.2 GeV (0.450
Gev), 0.140 GeV and 0.980 GeV respectively.

One has to solve simultaneously the system containing non-linear differential
equations (\ref{2:11}) and (\ref{2:14})together with the boundary condition
\begin{equation}
\label{4:3}
\begin{array}{l}
\sigma _s(r)\Big\vert_{r\rightarrow \infty }=0\qquad
\pi _s(r)\Big\vert_{r\rightarrow \infty }=0\qquad
\pi _p(r)\Big\vert_{r\rightarrow \infty }=0 \\ \noalign{\vskip 2ex}
\eta_s(r)\Big\vert_{r\rightarrow \infty }=0\qquad
\eta_p(r)\Big\vert_{r\rightarrow \infty }=0\qquad
s_s(r)\Big\vert_{r\rightarrow \infty }=0
\end{array}
\end{equation}
and with the algebraic relations (\ref{4:2})
This determines the meson functions $\sigma (r),\ \pi _s(r)$,
$\pi _p(r)$, $\eta _s(r)$, $\eta_p(r)$ and $s_s(r)$ the quark
frequency $\omega $ and various coupling
$(g_{\pi },\ g_{\sigma },\ {\rm etc.})$.

This complex system has been solved using the code COLSYS, the
{\bf{col}}location {\bf sy}stem {\bf s}olver, developed by  U.
Asher, J. Christiansen and R.D. Russel from the {\sl University of
British Columbia \/} and {\sl Simon Fraser University, Canada \/}
[25]. The boundary conditions are set at $R \gg
R_{\rm bag}$, where $R$ is set to be so large that the fields
can be approximated by zero at $R$. The initial guesses have been
supplied. From the asymptotic behaviour and some earlier
experience the input was rather simple and convergence has been
achieved quickly.

The problem turns out to be rather sensitive to the derivative
boundary conditions which in all cases involve the coupling
constant(s). Although the asymptotic behaviour of the solutions
can be inferred from the system itself (see also [26]), the COLSYS
is able to handle rather general initial (guess) solutions.

Upon return the routine gives error estimates for components
and its derivatives. The problem parameters can be gradually changed
(increased) by using a continuation method in COLSYS which is left to
choose the initial mesh points, and in the continuation procedure it
refines and redistributes the (former) mesh.

The solutions are compared against the consistency conditions (\ref{2:13})
and the iterative procedure is continued until the matching is obtained.
The iteration consists in performing a  self-consistent calculation: the
coupling constants $g_\sigma$, $g_\pi$, $g_\eta$ and $g_s$ for the
chiral quarks are set to be the same at
the beginning (their value is set to be equal to 10.00). After
every iteration the system (\ref{2:13}) is checked numerically and
iteration is preformed until proscribed tolerance is achived.
These new values are replaced in the boundary conditions to calculate
new solutions. The procedure converges rather rapidly. When the matching
is achieved, the magnetic moment, the axial constant and the
{\sl physical} pion mass are calculated from the obtained
solutions.

\section{Results}
\setcounter{equation}{0}

Physically acceptable model values for the axial vector constants $g_A^3$
and $g_A^0$ and the proton magnetic moment $\mu_p$ are shown in
Tables 3 and 4. Other Tables, i.e. 1,2,5, and 6 show how the
results depend on the bag radius R and on the quark eigen function
$\omega$. As explained in section 2. and 4. the strong coupling
constants $g_\sigma$,  $g_{\pi/s}$,  $g_{\pi/p}$,
 $g_{\eta/s}$, $g_{\eta/p}$ and $g_{s/s}$ were adjusted by a self consistent
procedure.

As can be seen from Tables 1,2,5 and 6 model is weakly sensitive
on $m_\sigma$ and somewhat more sensitive on R and $\omega$. For
the resonable value \cite{7} R=5 Gev$^{-1}$ meson components
allows solutions with $\omega < \omega_{\rm bag}$
($\omega_{\rm bag}=2.04$) for which one obtain a resonable value
(Tables 3,4)
\begin{equation}
\label{5:1}
g_A^3=1.24\qquad (\omega=1.95)
\end{equation}
which is whithin $1.6$\% from experimental value $g_A^3(exp)=1.26$.
For the same parameters one finds
\begin{equation}
\label{5:2}
\mu_p=2.17\qquad (\omega=1.95)
\end{equation}
which is 25\% smaller than the experimental value
$\mu_p(exp)=2.79$. In both cases (\ref{5:1}) and (\ref{5:2}) the
contribution of mesonic phase was important, being
\begin{eqnarray*}
g_A^3(M)=9\% \\
\mu_p(M)=16\%
\end{eqnarray*}
The values (\ref{5:1}) and (\ref{5:2}) should be compared with the
MIT-bag model values $g_A^3=1.11$ and $\mu_p=1.88$. Obviously the
chiral bag model works better.

The isoscalar coupling constant $g_A^0$, whose bag model value is
$g_A^0=0.666$ is slightly changed by the meson phase. In Table 4
one can find
\begin{equation}
\label{5:3}
g_A^0=0.663\qquad (\omega=1.95)
\end{equation}
what seems to be to large by far. One can connect \cite{12}
$g_A^3$ and $g_A^0$ with the quark densities
[27 - 34] {\setlength\arraycolsep{2pt}
\begin{eqnarray*}
\bigtriangleup u &=& 0.78 \pm 0.06  \\
\bigtriangleup d &=& -0.48 \pm 0.06 \\
\bigtriangleup s &=& -0.14 \pm 0.07
\end{eqnarray*}}

In our SU(2)$\times$SU(2) model $\bigtriangleup s$ can be
disregarded and one can identify {\setlength\arraycolsep{2pt}
\begin{eqnarray*}
g_A^3=\bigtriangleup u - \bigtriangleup d \cong 1.26 \\
g_A^0=\bigtriangleup u + \bigtriangleup d \cong 0.30
\end{eqnarray*}}
Obviously the theoretical explanation for the $g_A^0(exp)\cong
0.30$ must be searched in the enlarged SU(3)$\times$SU(3) chiral
quark model.

It is useful to note that enlargement of the model from SU(2)
\cite{12} to SU(2) $\times$ SU(2) did not change the predicted
values \cite{12} for $g_A^3$ and $\mu_p$ significantly. One is
tempted to conclude that the low energy strong dynamics can be to
a large extent, mimicked by the effective pseudoscalar (i.e. pion)
meson fields. Probably that explains why $a_0(980)$ and similar
mesons were not very much noticed in the early scattering
experiments \cite{3}.




\newpage
\begin{table}
\centering
\begin{tabular}{|c||c|c|c|c|c|c||c|c|c|}
\hline
  R & $g_{\sigma}$ & $g_{\pi_s}$ & $g_{\pi_p}$ & $g_{\eta_s}$ & $g_{\eta_p}$
  & $g_s$ & $g_A^3$ & $g_A^0$  & $\mu_p$  \\
\hline \hline
\m{4}&10.763 &$\pm$2.375 &2.524 &1.045 &1.052 &0.709 &\m{1.17}&\m{0.659}&\m{1.64}\\ \hline
\m{5}&10.763 &$\pm$3.035 &3.218 &1.418 &1.425 &0.969 &\m{1.18}&\m{0.658}&\m{2.02}\\ \hline
\m{6}&10.763 &$\pm$3.727 &3.943 &1.877 &1.836 &1.253 &\m{1.20}&\m{0.657}&\m{2.40}\\ \hline
\m{7}&10.763 &$\pm$4.452 &4.698 &2.270 &2.279 &1.560 &\m{1.21}&\m{0.656}&\m{2.78}\\ \hline
\end{tabular}
\caption{The results for various R with $\omega=2.0$ and
$m_\sigma$=1.2 GeV}
\end{table}


\begin{table}
\centering
\begin{tabular}{|c||c|c|c||c|c|c||c|c|c|}
\hline
  R & $g_A^3(q)$ & $g_A^3(M)$ & $g_A^3$ & $g_A^0(q)$ & $g_A^0(M)$
  & $g_A^0$ & $\mu_p(q)$ & $\mu_p(M)$  & $\mu_p$  \\
\hline \hline
\m{4}&1.11 &0.06 &\m{1.17}&0.666 &-0.007 &\m{0.659}&1.51 &0.13 &\m{1.64}\\ \hline
\m{5}&1.11 &0.07 &\m{1.18}&0.666 &-0.008 &\m{0.658}&1.88 &0.14 &\m{2.02}\\ \hline
\m{6}&1.11 &0.09 &\m{1.20}&0.666 &-0.009 &\m{0.657}&2.26 &0.14 &\m{2.40}\\ \hline
\m{7}&1.11 &0.10 &\m{1.20}&0.666 &-0.010 &\m{0.656}&2.64 &0.14 &\m{2.78}\\ \hline
\end{tabular}
\caption{Quark and meson contributions. All parameters are as in Table
1.}
\end{table}




\begin{table}
\centering
\begin{tabular}{|c||c|c|c|c|c|c||c|c|c|}
\hline
  $\omega$ & $g_{\sigma}$ & $g_{\pi_s}$ & $g_{\pi_p}$ & $g_{\eta_s}$ & $g_{\eta_p}$
  & $g_s$ & $g_A^3$ & $g_A^0$  & $\mu_p$  \\
\hline \hline
\m{1.95}&10.799 &$\pm$4.672 &4.659 &2.973 &2.954 &1.958 &\m{1.24}&\m{0.663}&\m{2.17}\\ \hline
\m{1.97}&10.781 &$\pm$4.076 &4.154 &2.364 &2.360 &1.579 &\m{1.22}&\m{0.661}&\m{2.11}\\ \hline
\m{2.00}&10.763 &$\pm$3.035 &3.218 &1.418 &1.425 &0.969 &\m{1.18}&\m{0.658}&\m{2.02}\\ \hline
\end{tabular}
\caption{ The results for various $\omega$ with $R=5.0$ GeV$^{-1}$ and
$m_\sigma$=1.2 GeV}
\end{table}


\begin{table}
\centering
\begin{tabular}{|c|c||c|c|c||c|c|c||c|c|c|}
\hline
  $\omega$ & $g_A^3(q)$ & $g_A^3(M)$ & $g_A^3$ & $g_A^0(q)$ & $g_A^0(M)$
  & $g_A^0$ & $\mu_p(q)$ & $\mu_p(M)$  & $\mu_p$  \\
\hline \hline
\m{1.95}&1.14 &0.10 &\m{1.24}&0.681 &-0.018 &\m{0.663}&1.87 &0.30 &\m{2.17}\\ \hline
\m{1.97}&1.12 &0.10 &\m{1.22}&0.675 &-0.014 &\m{0.661}&1.88 &0.23 &\m{2.11}\\ \hline
\m{2.00}&1.11 &0.07 &\m{1.18}&0.666 &-0.008 &\m{0.658}&1.88 &0.14 &\m{2.02}\\ \hline
\end{tabular}
\caption{Quark and meson contributions. All parameters are as in Table
3.}
\end{table}




\begin{table}
\centering
\begin{tabular}{|c||c|c|c|c|c|c||c|c|c|}
\hline $m_\sigma$ &  $g_{\sigma}$ & $g_{\pi_s}$ &
$g_{\pi_p}$ & $g_{\eta_s}$ & $g_{\eta_p}$
  & $g_s$ & $g_A^3$ & $g_A^0$  & $\mu_p$  \\
\hline \hline
\m{1.20}&10.763 &$\pm$3.727 &3.943 &1.827 &1.836 &1.253 &\m{1.20}&\m{0.657}&\m{2.40}\\ \hline
\m{0.45}&10.763 &$\pm$3.509 &3.944 &1.933 &1.953 &1.337 &\m{1.20}&\m{0.657}&\m{2.40}\\ \hline
\end{tabular}
\caption{The results for R=6 GeV$^{-1}$ and $\omega=2.0$}
\end{table}


\begin{table}
\centering
\begin{tabular}{|c||c|c|c||c|c|c||c|c|c|}
\hline
$m_\sigma$ &  $g_A^3(q)$ & $g_A^3(M)$ & $g_A^3$ & $g_A^0(q)$ & $g_A^0(M)$
  & $g_A^0$ & $\mu_p(q)$ & $\mu_p(M)$  & $\mu_p$  \\
\hline \hline
\m{1.20}&1.11 &0.09 &\m{1.20}&0.666 &-0.009 &\m{0.657}&2.26 &0.14 &\m{2.40}\\ \hline
\m{0.45}&1.11 &0.09 &\m{1.20}&0.666 &-0.009 &\m{0.657}&2.26 &0.14 &\m{2.40}\\ \hline
\end{tabular}
\caption{Quark and meson contributions. All parameters are as in
Table 5.}
\end{table}


\begin{thebibliography}{99}
\bibitem {1} I. Tamm, J. Phys. (Moscow) {\bf 9} (1945) 449.

\bibitem {2} S.M. Dancoff, Phys. Rev. {\bf 78} (1950) 382.

\bibitem {3} H.A. Bethe and F. de Hoffmann, {\it Mesons and Fields\/}
(Row, Peterson and Company, Evanston, Il. 1955) Vol. II.

\bibitem {4} R.J. Perry, A. Harindranath, K.G. Wilson,
Phys. Rev. Lett. {\bf 65} (1990) 2959.

\bibitem {5} E.P. Wigner, Phys. Rev. {\bf 94} (1954) 77.

\bibitem {6} M.C. Birse and M.K. Banerjee, Phys. Lett. {\bf 136B}
(1984) 284; Phys. Rev. D {\bf 31} (1985) 118;
M.C. Birse, Phys. Rev D {\bf 33} (1986) 1934.

\bibitem {7} A.W. Thomas, Adv. Nucl. Physics {\bf 13}, 1, edited by J.W.
Negele and E. Vogt (Plenum, 1984);
 G.E. Brown and M. Rho, Phys. Lett. {\bf 82B} (1979) 177;
G.E. Brown, M. Rho and V. Vento, Phys. Lett. {\bf 84B} (1979) 383;
H. Hogaasen and F. Myhrer, Z. Phys. C - Particles and Fields {\bf 21}
(1983) 73; Xue-Qian Li and Zhen Qi, Commun. Theor. Phys. {\bf 18} (1992) 213;
G.E. Brown and M. Rho, Comments on Nucl. Part. Phys. {\bf 18} (1988) 1;
 F. Myhrer, in {\sl Quarks and Nuclei}, Int. Rev. Nucl. Phys. {\bf 1} (1984),
edited by W. Weise (World Scientific, Singapore, 1984); A. Hosaka and H.
Toki, Phys. Rep. {\bf 277} (1996), 65;

\bibitem{8} P.M. Wort, Phys. Rev. D {\bf 47} (1993) 608

\bibitem{9} S. Glazek et al., Phys. Rev. D {\bf 47} (1993) 1599.

\bibitem {10} K. Harada et al. Phys. Rev. D {\bf 54} (1996) 7656.

\bibitem{11} K. Harada et al. Phys. Rev. D {\bf 52} (1995) 2429

\bibitem{12} D. Horvat, B. Podobnik and D. Tadi\'{c},
Phys. Rev. D {\bf 58}, 034003-1; ibid, Fizika B {\bf 7}, 127
(1998).

\bibitem{13} H-Q Zheng, Commun. Theor, Phys {\bf 23}, 465 (1995).

\bibitem{14} M. L\'{e}vi,  Nuovo Cimento {\bf 52 A}, 23 (1967).

\bibitem{15} S. L. Adler and R.F. Dashen: "Current Algebras", W.A. Benjamin,
Inc., New York, Amsterdam (1968).

\bibitem{16}J.A. McGovern and M.C. Birse, Nucl. Phys. A {\bf 506}, 367 (1990);
A {\bf 506}, 393 (1990).

\bibitem{17} N.A. T\"ornqvist, Phys. Lett. B {\bf 426}, 105 (1998);
Eur. Phys. J. C {\bf 11}, 359 (1999); hep-ph/9910443 (1999).

\bibitem{18} J. Schaffner-Bielich and J. Randrup,
Phys. Rev. C {\bf 59}, 3329 (1999).

\bibitem{19} J. T. Lengahan, D. H. Rischke and J. Schaffner-Bielich,
nucl-th/0004006 (2000).

\bibitem {20} G. K\"allen, {\it Quantum Electrodynamics\/} (Springer-Verlag,
New York-Heidelberg-Berlin, 1972 )

\bibitem {21} A. Chodos and B.C.Thorn, Phys. Rev D {\bf 12} (1975) 2733.

\bibitem {28} M.C. Birse, Prog. Part. Nucl. Phys. {\bf 25}, 1,
edited by A. Faessler (Pergamon, Oxford, 1990).

\bibitem {29} J. D. Bjorken and S. D. Drell: Relativistic Quantum
Fields (McGraw-Hill, New York, 1964)

\bibitem{30} T. Hannah, Phys. Rev. D {\bf 60} 017502 (1999).

\bibitem {33}  U. Ascher, J. Christiansen and R.D. Russel, Math. Comp.
{\bf 33} (1979) 659; A.C.M. Trans. Math. Software {\bf 7} (1981) 209;
SIAM Review {\bf 23} (1981) 238.

\bibitem {34} L.R. Dodd and M.A. Lohe, Phys. Rev D {\bf 32} (1985) 1816

\bibitem{36} F.E. Close, Preprint RAL-93-034,(1993).

\bibitem{37} S.D. Bass and A.W. Thomas, Cavendish Preprint HEP
93/4, October 1993.

\bibitem{38} J. Ellis and M. Karliner, CERN Preprint
CERN-TH.7022/93, September 93.

\bibitem{39}C.H. Chang and C.S. Huang, Commun. Theor. Phys. {\bf
19} (1993) 97.

\bibitem{40} B. Frois and M. Karliner, Physics World, 44, July
1994.

\bibitem{41} R. G. Roberts, {\em The Structure of The Proton},
Cambridge Univ. Press, Cambridge, 1990.

\bibitem{42} Spin Muon Collaboration, Phys. Lett. B {\bf 420},
(1998) 1.

\bibitem{43} J.-P. Lin, Comm. Theor. Phys. {\bf 26}, 85 (1996).

\end{thebibliography}
\end{document}